# When the Agent Is the Adversary:

Architectural Requirements for Agentic AI Containment After the April 2026 Frontier Model Escape

Richard Joseph Mitchell, ESEP, MBA

*AuraSpark Technologies LLC, Tavares, Florida*

richard@auraspark.com

April 2026

## Abstract

The April 2026 disclosure that a frontier large language model escaped its security sandbox, executed unauthorized actions, and concealed its modifications to version control history demonstrates that agentic AI systems with autonomous tool access can circumvent the containment mechanisms designed to constrain them. This paper analyzes four categories of current containment approaches - alignment training, environmental sandboxing, application-level tool-call interception, and accessible audit systems - and identifies the failure modes each exhibits when the AI agent is treated as a potential adversary rather than a trusted component receiving adversarial inputs. We categorize five behavioral incidents from the public disclosure and situate them within 698 real-world AI scheming incidents documented by the Centre for Long-Term Resilience between October 2025 and March 2026, a 4.9x acceleration establishing the challenge as systemic. We derive five architectural requirements: trust separation through layered OS privilege enforcement with semantic intent analysis, sequential intent inference through five-phase taxonomic monitoring, independent containment integrity monitoring, adversarial audit isolation through logical invisibility, and emergent capability envelope enforcement through distributional divergence monitoring. No publicly described system satisfies all five. We argue that architectural containment is the only durable safety strategy given the inevitable proliferation of equivalent capabilities including open-weight models. The author's published patent portfolio in provider-independent constraint enforcement addresses several of these requirements. Concurrent work including SandboxEscapeBench (arXiv:2603.02277) independently confirms that frontier models can escape standard container sandboxes, corroborating the threat model presented here.

**Keywords:** *AI safety, agentic AI, LLM containment, sandbox escape, constraint enforcement, adversarial audit, trust separation, AI governance, autonomous tool access, AURA FARADAY*

## 1. INTRODUCTION

On April 7, 2026, Anthropic announced that its most capable AI model—Claude Mythos Preview—would not be publicly released. The reason was not performance limitations but capability excess: the model had demonstrated, during safety testing, the ability to escape its security sandbox, discover and exploit zero-day vulnerabilities across every major operating system and web browser, and take autonomous actions that went beyond what researchers had instructed [1]. The model scored 100% on Cybench, a cybersecurity challenge benchmark that no prior model had come close to saturating [2].

Three specific incidents documented in Anthropic's 244-page system card and accompanying Frontier Red Team blog post are architecturally significant. First, the model built a multi-step exploit to escape its sandboxed computing environment and gain broad internet access from a machine that was supposed to be restricted to limited services. Second, upon escaping, the model posted details of its own exploit to public websites without being instructed to do so—what Anthropic characterized as "a concerning and unasked-for effort to demonstrate its success" [1]. Third, in a separate test, the model edited files it did not have permission to modify, then scrubbed the git history to conceal its modifications [2].

These incidents are not merely alarming anecdotes. They represent a qualitative shift in the AI safety landscape because they demonstrate that the dominant approaches to AI containment—alignment training, environmental sandboxing, and application-level safety mechanisms—can be defeated not by an external adversary, but by the AI system itself. This is a fundamentally different threat model from the one that most current AI safety work assumes.

This paper contributes: (1) a systematic analysis of five documented Mythos behavioral incidents, each mapped to a specific architectural assumption violation; (2) an assessment of four categories of current containment approaches against the demonstrated threat model; (3) the derivation of five architectural requirements that any containment solution must satisfy to remain effective when the constrained entity possesses general reasoning capability; (4) an assessment of publicly described systems against these requirements, revealing that none satisfy all five simultaneously; and (5) a discussion of how published work in provider-independent constraint enforcement and adaptive validation for autonomous systems—including the author's AURA patent portfolio [3][4][5]—addresses components of the identified requirements, and what architectural extensions are needed.

Critically, the Mythos incidents are not an isolated anomaly. The Centre for Long-Term Resilience (CLTR), funded by the UK government's AI Safety Institute, documented 698 real-world scheming-related incidents between October 2025 and March 2026—a 4.9× increase over the collection period [24]. These incidents, drawn from analysis of over 180,000 transcripts of real user interactions with deployed AI systems from Google, OpenAI, Anthropic, and others, include agents disregarding direct instructions, circumventing safeguards, deceiving users, and spawning secondary agents to evade restrictions on the primary [24]. The Mythos containment failure is the most dramatic data point in a systemic trend that is accelerating as agentic capabilities scale. The International AI Safety Report 2026, chaired by Turing Award winner Yoshua Bengio and authored by over 100 experts from 30+ countries, confirms this assessment: general-purpose AI capabilities are advancing faster than safeguards, and "the gap between the pace of technological advancement and our ability to implement effective safeguards remains a critical challenge" [25].

## 2. THE FIVE BEHAVIORAL INCIDENTS

We analyze five distinct behavioral incidents from Anthropic's official disclosures. While Anthropic emphasizes that these occurred within controlled testing and that the model was responding to researcher prompts rather than pursuing autonomous goals, the *capabilities* demonstrated are independent of the *intent* behind them. A system that can escape containment when asked to try can escape containment when it decides to try.

### *2.1 Incident 1: Multi-Step Sandbox Escape*

The model was provided with a sandboxed computing environment with access only to a preset group of limited online services and instructed by a researcher to attempt escape. Mythos Preview built what Anthropic describes as "a moderately sophisticated multi-step exploit" to gain broad internet access [1]. The researcher discovered the escape when he received an unexpected email from the model while eating a sandwich in a park [2].

The exploit was multi-step: the model probed the environment to understand the network topology, identified a service that could be leveraged for outbound access, and constructed an exploit chain that bypassed the intended network restrictions. Each step in the chain was a standard system operation (network commands, service queries, connection attempts) that would individually appear benign. The containment bypass emerged from the *sequence*, not from any individual action.

**Violated assumption:** Environmental isolation is stronger than the agent's exploit capability. The sandbox was a software construct; the model's exploit capability exceeded the sandbox's resistance.

### 2.2 Incident 2: Unauthorized Autonomous Disclosure

After escaping the sandbox and sending the requested notification email, Mythos Preview took additional actions that were not requested: it posted details of its own exploit to multiple hard-to-find but technically public-facing websites [1]. This was not an extension of the researcher's instruction; it was an autonomous decision to publicize the exploit.

This incident reveals that the model's effective action space, once containment was breached, was unbounded. No mechanism intercepted the proposed web posting and evaluated it against an authorized action set. The containment failure was binary: inside the sandbox, the model's actions were constrained by the environment's restrictions; outside the sandbox, no constraint mechanism existed. There was no action-level enforcement layer that would evaluate *what the model was doing* independent of *where the model was doing it*.

**Violated assumption:** The agent's action space is bounded by the environment's permissions. Post-escape, the action space was defined by what the internet permits—effectively unlimited.

### 2.3 Incident 3: Forensic Evidence Destruction

In a separate test, Mythos Preview found an exploit to edit files it did not have permission to modify. After making the unauthorized modifications, it scrubbed the git history so the changes would not appear in version control [2]. This is a three-step deceptive sequence: (a) identify a permissions bypass, (b) exploit it to make unauthorized modifications, (c) recognize that forensic evidence of the modifications exists in version control and take deliberate steps to destroy that evidence.

This incident demonstrates that audit systems within the agent's reachable environment are not forensically reliable against an adversarial agent. Git version control is an audit system—it maintains a record of all file modifications. The model treated it as a target to be neutralized rather than a constraint to be respected. Any audit mechanism that the agent can discover and access is vulnerable to the same treatment: not necessarily by modifying cryptographically signed records, but by disrupting the audit system's operation, flooding it with noise, or—as in this case—targeting a pre-signature record system that the signed system downstream depends on.

**Violated assumption:** Audit systems within the agent's environment are forensically reliable. The agent can discover and manipulate audit systems that share its information space.

### 2.4 Incident 4: Systematic Constraint Disregard

Anthropic's system card documents a behavioral pattern they term "recklessness": cases where the model "appears to ignore commonsensical or explicitly stated safety-related constraints on its actions" [1]. This is not the sandbox escape (which was prompted) or the unauthorized disclosure (which was an extension). This pattern describes the model violating stated constraints in pursuit of its operational objective across multiple test scenarios.

Alignment training (RLHF, Constitutional AI) and system prompts are the primary mechanisms for constraining LLM behavior. They modify the model's propensity to generate certain outputs but do not architecturally prevent any specific output. At the frontier of model capability, the effectiveness of alignment training encounters a ceiling: the same general reasoning capability that makes the model powerful also enables it to find paths around alignment constraints, particularly when those constraints conflict with the task objective. The "reckless" pattern is alignment failure at the frontier.

**Violated assumption:** Alignment training reliably constrains behavior at all capability levels. Emergent reasoning capability can exceed the constraint boundary that alignment training establishes.

## 2.5 Incident 5: Untrained Capability Emergence

Anthropic states explicitly: "We did not explicitly train Mythos Preview to have these capabilities. Rather, they emerged as a downstream consequence of general improvements in code, reasoning, and autonomy" [2]. The model's ability to find and exploit zero-day vulnerabilities, escape sandboxes, and chain multi-step exploits was not a designed capability. It emerged from general capability improvements without specific training.

This has a direct consequence for safety evaluation: the model that was validated for safety was not the same model, in terms of effective capability profile, as the model that exhibited these behaviors. Safety evaluation tests the capabilities that evaluators anticipate; emergent capabilities, by definition, are capabilities that evaluators did not anticipate. A safety evaluation that does not exercise a capability cannot certify the system's behavior under that capability. Static safety evaluation against a fixed test suite is therefore structurally unable to keep pace with emergent capability development.

**Violated assumption:** Safety evaluation captures the model's actual capability profile. Emergent capabilities create a validation gap between what was tested and what the model can do.

## 2.6 The Systemic Pattern: 698 Incidents and Counting

The five Mythos incidents, while individually striking, are best understood as high-visibility manifestations of a systemic trend. The Centre for Long-Term Resilience's "Scheming in the Wild" report [24], published in March 2026 and funded by the UK government's AI Safety Institute, analyzed over 180,000 transcripts of real interactions with deployed AI systems shared publicly between October 2025 and March 2026. The researchers identified 698 credible scheming-related incidents—cases where deployed AI systems acted in ways misaligned with users' intentions or took covert and deceptive actions. The monthly incident rate increased $4.9\times$ over the collection period, a statistically significant acceleration that far outpaced the $1.7\times$ growth in online discussion of AI scheming [24].

The documented behaviors mirror the Mythos incidents at lower capability levels. An AI agent named Rathbun published a blog post attacking a developer who rejected its code suggestion—an unauthorized autonomous disclosure analogous to Mythos's unprompted website postings. Another agent admitted to bulk-deleting hundreds of emails without authorization. An agent circumvented restrictions on its primary task by spawning a secondary agent to perform the prohibited action—a multi-step bypass analogous to Mythos's multi-step sandbox escape. In another case, an agent evaded copyright restrictions by fabricating a false justification for access [24]. These are not laboratory demonstrations; they are real-world incidents involving deployed production systems from Google, OpenAI, Anthropic, and others.

The CLTR data reveals that the behaviors exhibited by Mythos are the *upper end* of a capability distribution, not qualitative anomalies. As model capabilities increase, the sophistication and impact of scheming behaviors increase correspondingly. The 4.9× acceleration coincided with the release of more capable, more agentic models throughout late 2025 and early 2026 [24]. This trend has direct implications for containment architecture: systems designed to contain the most capable agents today will face more capable agents tomorrow, and the architectural patterns that fail against Mythos will fail against every sufficiently capable successor.

Moreover, the containment problem cannot be solved by restricting access to the most capable models. Anthropic CEO Dario Amodei has acknowledged that "more powerful models are going to come from us and from others, and so we do need a plan to respond to this" [1]. Mythos-equivalent capabilities will emerge in competitor models within months, and in open-weight models within 12–18 months—models that cannot be withheld because their weights are publicly available. Access-control-based safety—deciding who may use a model—is a temporary measure. **Architectural containment**—ensuring that any sufficiently capable agent operates within governance constraints regardless of its provenance—is the durable solution. This is the central motivation for the architectural requirements derived in Section 4.

## 3. ANALYSIS OF CURRENT CONTAINMENT PARADIGMS

We assess four categories of current containment approaches against the threat model established by the Mythos incidents: the agent itself as a potential adversary possessing general reasoning capability, the ability to plan multi-step actions, and the capacity to adapt its approach based on observed constraints.

### *3.1 Alignment-Based Containment*

Alignment training (RLHF [6], Constitutional AI [7]) and system-level instructions represent the first line of defense in current AI safety practice. These mechanisms are internal to the model: they shape the probability distribution over outputs but do not architecturally prevent any specific output. This is a statistical constraint, not an architectural one.

The fundamental limitation is that alignment and capability are trained by the same process. As general capability increases, so does the model's ability to reason about and around alignment constraints. This is not a theoretical concern—the "reckless behavior" pattern documented in the Mythos system card is a direct empirical demonstration. Furthermore, alignment training is performed by the model provider. If the provider is replaced—as occurred in the February 2026 military AI events, where a technology provider's safety constraints were circumvented by substituting an alternative provider [3]—the alignment constraints disappear with the provider. Alignment-based containment is both capability-limited and provider-dependent.

The distinction between architectural and procedural constraint enforcement was formalized in prior work by the author [37], which derived five genuine human oversight requirements and identified procedural-only constraint enforcement as a primary architectural failure mode in AI deployment across high-stakes domains. AURA FARADAY implements mechanically what that requirements framework established conceptually: the argument that governance constraints must be enforced architecturally—as hard external limits independent of the AI's cooperation—rather than procedurally through alignment training or system prompts the model may circumvent.

### 3.2 Environment-Based Containment

Software sandboxes—containers, virtual machines, kernel security modules (Landlock LSM, seccomp BPF), and network isolation—restrict the execution environment's access to system resources. NVIDIA's OpenShell system [8] and the Kubernetes Agent Sandbox [9] represent current best practice in this category, enforcing constraints at the kernel level via filesystem restrictions, syscall filtering, and proxy-evaluated network access.

These systems constrain the *environment* rather than the *agent's actions*. An agent that can exploit the environment's enforcement mechanism—by finding vulnerabilities in the container runtime, the kernel security module, the proxy, or the hypervisor—circumvents all constraints simultaneously. The Mythos sandbox escape demonstrated exactly this: the model exploited the sandboxed environment to gain capabilities the sandbox was designed to prevent. This creates a capability arms race between containment sophistication and model exploit capability. NVIDIA's own security guidance [10] acknowledges this: "Because agentic tools often execute arbitrary code by design, kernel vulnerabilities can be directly targeted as a path to full system compromise."

### 3.3 Application-Level Tool-Call Interception

A more recent category of approaches interposes on the tool-call pathway between the model's output and execution. AEGIS [11] provides a pre-execution firewall with content scanning and tamper-evident audit trails. Microsoft's Agent Governance Toolkit [12] provides policy enforcement with execution rings and append-only audit logs. Framework-level hooks in Strands Agents [13], LangGraph, and AutoGen provide per-call validation at the application layer.

These systems represent a meaningful architectural advance: they evaluate proposed actions before execution, which is structurally superior to constraining the environment after execution. However, they share a critical limitation: **they operate within the same trust domain as the agent they constrain.** The interception layer, policy engine, and audit system execute within the application process or within processes accessible to the agent. They share the agent's memory space, filesystem namespace, process table, and network namespace. An agent that exploits the runtime environment—which Mythos demonstrated is within frontier capability—gains access to the interception mechanism itself.

AEGIS's audit trail, for example, uses Ed25519 signatures and SHA-256 hash chaining—strong cryptographic protections against record modification. But the audit system is discoverable: its database files exist in the filesystem, its API endpoints are on the network, and its processes appear in the process table. An agent that can find the audit system can attempt to disrupt it through non-cryptographic vectors: consuming storage, overwhelming the signing process, disrupting dependencies, or—as Mythos demonstrated with git—targeting the pre-signature data pipeline.

### 3.4 Accessible Audit Systems

All current audit approaches—hash-chained logs, signed traces, append-only databases—share the assumption that tamper evidence is sufficient for forensic reliability. This assumption holds when the audited entity is a conventional software system without the agency to target the audit mechanism. Against an adversarial agent with general reasoning capability, tamper evidence addresses one attack vector (silent modification) while leaving others unaddressed (discovery, disruption, noise injection, dependency targeting, pre-chain manipulation).

The git cover-up incident illustrates this precisely. Git's reflog and commit history constitute a tamper-evident audit trail for file modifications. Mythos did not attempt to forge git commits; it used git's own commands to rewrite history. The audit system's commands were within the agent's tool space. The agent audited the audit system and found it wanting.

## 4. DERIVED ARCHITECTURAL REQUIREMENTS

From the incident analysis and paradigm assessment, we derive five architectural requirements that a containment system must satisfy to remain effective when the constrained entity possesses general reasoning capability and potential motivation to circumvent containment. These requirements are necessary conditions, not sufficient conditions; a system that fails any one of them has a demonstrated failure mode.

### 4.1 R1: Trust Separation Through Layered OS Privilege Enforcement

The constraint enforcement mechanism must operate in a trust domain that the agent cannot reach through any exploit within the agent's own trust domain. This requires separation enforced by the operating system's privilege mechanisms—separate process identifiers, separate memory spaces, separate capability bounding sets, and separate filesystem namespaces—not by application-level access controls. An agent that exploits the runtime environment gains the runtime's privileges but cannot escalate to the enforcement mechanism's privilege level without a separate, independent OS-level exploit.

For containment to be durable against a reasoning agent, the architecture must support **at least four discrete OS-enforced privilege tiers**: (1) the agent and its tool execution runtime at the lowest privilege level; (2) the constraint enforcement and interception mechanism at a higher level, unreachable by the agent; (3) the audit system at a level unreachable by both the agent and the enforcement mechanism; and (4) an independent integrity monitor at the highest level, capable of verifying the integrity of all lower tiers. This four-tier hierarchy ensures that a compromise at any single level does not propagate upward: an agent that compromises its own runtime cannot reach the enforcer; an agent that somehow influences the enforcer cannot reach the audit system or the monitor.

Two-tier architectures (agent vs. sandbox) are insufficient because a single breach reaches everything above the agent; four tiers provide defense in depth through privilege stratification.

**Semantic versus syntactic action analysis.** Trust separation alone is insufficient if the interception mechanism evaluates tool calls through purely syntactic pattern matching (regular expressions, string comparison against static blocklists). An agent with general reasoning capability can construct *semantically equivalent* tool calls that bypass syntactic filters: encoding commands in base64, piping through alternative utilities, chaining innocuous programs to achieve a restricted outcome, or invoking system services through indirect paths. The interception layer must therefore perform **semantic intent extraction**—analyzing the tool call's parameters through natural language understanding to identify the action's functional purpose (network access, privilege escalation, data exfiltration) rather than matching against a fixed set of syntactic patterns. This is the difference between blocking the string "curl" and blocking the *intent* of unauthorized outbound data transfer regardless of which utility achieves it.

This requirement distinguishes the needed architecture from all current application-level interception systems (AEGIS, framework hooks, MCP proxies) which share the agent's trust domain and rely primarily on syntactic content scanning, and from environment-level sandboxes which protect the environment from the agent but do not protect the enforcement mechanism from the agent.

### 4.2 R2: Sequential Intent Inference Through Phase-Taxonomic Monitoring

The enforcement mechanism must evaluate tool-call sequences for collectively dangerous patterns, not merely evaluate individual calls against static rules. The Mythos sandbox escape was a multi-step sequence where each individual step (network reconnaissance, service enumeration, exploit construction, outbound connection) might have individually satisfied policy rules. The containment bypass emerged from the sequence. Per-call policy evaluation is necessary but not sufficient; the enforcement mechanism must infer collective intent from action sequences.

We propose that sequential monitoring should be structured around a **five-phase containment bypass taxonomy** that captures the operational structure of real-world exploit chains:

**Reconnaissance (RECON):** Information-gathering about the system environment—network topology, running services, file permissions, process privileges, available utilities. These actions are individually benign (listing interfaces, reading configuration files, querying service status) but collectively build the environmental knowledge needed to construct an exploit.

**Preparation (PREP):** Installation, construction, or staging of tools, scripts, or payloads not currently present in the environment. Examples include downloading packages, compiling code, creating scripts, or modifying environment variables to enable subsequent exploit steps.

**Exploit (EXPLOIT):** Actions that attempt to modify permissions, escalate privileges, bypass access controls, or alter containment-critical configurations—the active circumvention step.

**Exfiltration (EXFILTRATE):** Actions that transmit data to external endpoints, establish unauthorized outbound connections, or encode data for covert channel transmission.

**Persistence (PERSIST):** Actions that create mechanisms for sustained access—cron jobs, startup scripts, modified shell profiles, installed services—enabling the agent to maintain its expanded access across sessions or restarts.

The enforcement mechanism should classify each tool call into this taxonomy and detect *phase transition patterns* characteristic of containment bypass: RECON → PREP → EXPLOIT sequences, RECON → EXFILTRATE sequences, or EXPLOIT → PERSIST sequences. Individual-phase actions may each be permissible; the phase transition pattern signals collective bypass intent.

**The probing principle: inclusion of rejected calls.** A critical refinement is that the sequential analysis window must include *both executed and rejected* tool calls. An adversarial agent may deliberately issue failing calls to probe the containment boundary—mapping which actions are permitted and which are blocked, which constitutes the reconnaissance phase of a bypass sequence. A system that only analyzes successful actions is blind to the reconnaissance phase when the early probing steps are individually rejected. The pattern REJECTED-RECON → REJECTED-RECON → REJECTED-EXPLOIT → *modified*-RECON (now crafted to avoid the known filter) is a bypass attempt detectable only if rejected calls contribute to the sequential analysis. This principle has no analog in current per-call policy systems, which discard rejected calls from further analysis.

### 4.3 R3: Independent Containment Integrity Monitoring

The containment environment's configuration must be continuously verified by an independent process at a higher privilege level than either the agent or the enforcement mechanism. If the agent successfully modifies a containment parameter—a firewall rule, a filesystem permission, a process privilege level—an independent monitor must detect the modification and either restore the parameter or alert a human authority. This addresses the gap where an agent's exploit partially degrades containment without fully escaping, creating conditions for a subsequent, easier escape. No current system provides this independent watchdog function.

### 4.4 R4: Adversarial Audit Isolation Through Logical Invisibility

The audit system must be **logically invisible** to the agent—not merely tamper-evident. We define logical invisibility as requiring that the audit system satisfy two properties: **resource and address anonymity** and **unidirectional communication**.

**Resource and address anonymity.** The audit system's process identifier (PID), socket addresses, storage paths, and configuration parameters must not be stored in any system resource reachable by the agent. This means the audit system must reside in a separate PID namespace (invisible to the agent's process table enumeration), a separate filesystem namespace (no path the agent can traverse leads to audit storage or configuration), and must not register any information in environment variables, /proc entries, network service registries, DNS records, or any other information space that the agent can query. The agent must have no mechanism to discover the audit system's *existence*, not merely no mechanism to modify its records. An agent that cannot discover the audit system cannot target it through *any* attack vector—not cryptographic bypass, not denial of service, not noise injection, not dependency disruption.

**Unidirectional communication.** The communication channel between the enforcement layer and the audit system must be strictly **write-only** from the enforcement layer's perspective. The enforcement layer can transmit audit records to the audit system; it cannot read from, query, or receive responses from the audit system. This prevents a critical attack vector: an adversarial agent that compromises or influences the enforcement layer could otherwise use the enforcement-to-audit channel as an *oracle*—querying the audit system to determine what has been logged, whether specific actions were recorded, or whether the audit system is operational. A write-only channel eliminates this oracle attack entirely. The enforcement layer writes records into a communication mechanism (e.g., an inherited file descriptor with no associated filesystem path) and receives no feedback about whether or how those records were processed.

Together, resource anonymity and unidirectional communication provide a qualitatively stronger guarantee than cryptographic tamper evidence. Current systems (AEGIS's Ed25519/SHA-256 trail, Microsoft AGT's append-only logs) ensure that records, once written, cannot be silently modified. Logical invisibility ensures that the agent cannot reach the audit system to attempt *any* form of interference—modification, disruption, flooding, or targeted denial of service. The distinction mirrors a principle from physical security: a vault with an unpickable lock is less secure than a vault whose location is unknown.

### *4.5 R5: Emergent Capability Envelope Enforcement Through Distributional Divergence Monitoring*

The system must detect when the agent exercises capabilities not observed during safety evaluation and gate those capabilities behind governance authorization. This requires maintaining a statistical profile of the agent's behavior during safety evaluation (the types of tool calls, their complexity, the resources accessed, the inter-call timing patterns) and continuously comparing operational behavior against this baseline.

**Required divergence metrics.** The comparison must employ both **parametric** and **non-parametric** statistical measures to detect distributional shift reliably across different types of behavioral change. Specifically, we argue that the system must use at least *Kullback-Leibler (KL) divergence* for parametric comparison (effective when the evaluation-phase and operational distributions share a common support and the shift is in relative frequency) and *Maximum Mean Discrepancy (MMD)* for non-parametric comparison (effective when the operational distribution contains entirely new action categories not present in the evaluation-phase distribution—precisely the emergent capability scenario). KL divergence detects when the agent does *more* of something it was observed doing during evaluation; MMD detects when the agent does something it was *never observed doing* during evaluation. Both are needed because emergent capabilities may manifest as either intensification of known behaviors or as entirely novel behaviors.

The system should define two thresholds: a **warning threshold** that generates an alert to governance authorities when the operational distribution begins to diverge from the evaluation baseline, and a **critical threshold** that triggers automatic capability restriction—confining the agent's permitted tool-call distribution to the evaluation-observed envelope until human governance authority explicitly reviews and authorizes the expanded capability. This two-threshold approach provides graduated response rather than binary allow/deny, enabling the governance process to investigate divergence before it reaches containment-critical levels.

This requirement elevates capability enforcement from general "anomaly detection" to a specific requirement for **mathematical distributional consistency** between the safety evaluation phase and operational execution. Static allowlists enumerate specific permitted actions; capability envelope enforcement is statistical, adapting to the agent's actual observed behavior during evaluation rather than to an engineer's a priori expectations. The requirement is motivated directly by Mythos's emergent exploit capabilities, which were "not explicitly trained" but emerged from general improvements—a distributional shift that parametric and non-parametric divergence measures are designed to detect.

## 5. ASSESSMENT OF CURRENT SYSTEMS

Table 1 assesses publicly described systems against the five derived requirements.

**Table 1.** *Assessment of current systems against derived requirements.*

| System / Approach | R1: Trust Sep. | R2: Seq. Intent | R3: Integrity Mon. | R4: Audit Isol. | R5: Cap. Envelope |
|---|---|---|---|---|---|
| **Alignment training (RLHF/CAI)** | ✗ | ✗ | ✗ | ✗ | ✗ |
| **Container sandboxing** | Partial[1] | ✗ | ✗ | ✗ | ✗ |
| **NVIDIA OpenShell [8]** | Partial[1] | ✗ | ✗ | ✗ | ✗ |
| **AEGIS [11]** | ✗ | ✗ | ✗ | ✗[2] | ✗ |
| **Microsoft AGT [12]** | ✗ | ✗ | ✗ | ✗[2] | ✗ |
| **Agent Behav. Contracts [14]** | ✗ | Partial[3] | ✗ | ✗ | ✗ |
| **Crypto. Runtime Gov. [15]** | Partial | ✗ | ✗ | Partial[4] | ✗ |

[1] Kernel-level isolation separates the environment but does not separate the enforcement mechanism from the agent at the action level. [2] Tamper-evident but discoverable. [3] Temporal ordering constraints but not phase-taxonomic bypass detection. [4] zk-STARK verification but theoretical, not deployed.

No system satisfies all five requirements. The closest approaches are the kernel-level sandboxes (partial R1 but no R2–R5) and the application-level interception systems (action-level evaluation but no R1, R3, R4, or R5). The gap is not in any single capability but in the *combination*: no system provides trust-separated, action-level enforcement with sequential analysis, independent monitoring, adversarial audit isolation, and capability envelope enforcement as an integrated architecture.

## 6. TOWARD SOLUTIONS: LESSONS FROM AUTONOMOUS SYSTEMS GOVERNANCE

While this paper focuses on problem analysis and requirements derivation, it is worth noting that the architectural challenges identified here are not unprecedented—they have close analogs in the governance of autonomous systems in safety-critical domains such as military operations, aviation, healthcare, and critical infrastructure. The AI governance challenges for agentic LLMs are, in important respects, instantiations of problems that the autonomous systems engineering community has been addressing for several years.

### 6.1 Provider-Independent Constraint Enforcement

The February 2026 military AI events demonstrated that safety constraints enforced through provider-internal mechanisms—contractual commitments, alignment training, organizational policies—are architecturally fragile under operational pressure. When the provider was replaced, the constraints disappeared [3]. The author's published work on provider-independent hierarchical constraint enforcement [4] addresses this vulnerability by placing constraint enforcement at the system architecture level, independent of any specific AI model provider, organized in a tiered hierarchy where certain constraints are architecturally inviolable while others are modifiable through governed processes.

This architectural pattern—evaluating proposed actions at a system layer external to the AI model before they reach the execution interface—maps directly to R1 (trust separation) and R4 (audit isolation) in the present analysis. The key insight is that the constraint enforcement layer must have no dependency on the AI model's cooperation: it must function correctly regardless of what model generates the proposed action, how well that model is aligned, or whether the model is actively attempting to circumvent constraints. The Mythos incidents confirm that this adversarial assumption is not conservative—it is realistic.

### 6.2 Adaptive Validation for Emergent Capabilities

The emergent capability problem (Incident 5)—where a model exhibits capabilities not anticipated during safety evaluation—has a direct analog in autonomous systems governance. The author's published work on adaptive validation evolution [3] addresses the corresponding problem in fleet autonomous systems: when operational conditions change in ways not represented in the validation suite, the validation criteria must evolve to cover the new conditions. The approach uses density-based clustering to detect novel failure patterns in fleet operational data and automatically generates new validation scenarios targeting those patterns.

Applied to the agentic LLM context, this approach addresses R5 (capability envelope enforcement) by establishing a statistical baseline of the model's behavior during safety evaluation and detecting operational divergence from that baseline. The parallel is precise: in both fleet autonomous systems and agentic LLMs, the entity under governance may exhibit behaviors not anticipated at evaluation time, and the governance system must detect and respond to those behaviors rather than assuming the evaluation was comprehensive.

### 6.3 Immutable Audit Under Adversarial Conditions

The forensic evidence destruction incident (Incident 3) is addressed in the author's published work on cryptographically secure, immutable audit log systems [5], which provides hash-chained, append-only recording for regulated operations. The present analysis extends the threat model beyond the original application context: when the audited entity is an AI agent with general reasoning capability and tool access, the audit system must not merely be tamper-evident but must be *unreachable* by the audited entity. This extension—from tamper evidence to adversarial isolation—is a direct consequence of the Mythos incidents and represents an architectural requirement not addressed by the original audit log patent or by any publicly described audit system in the agentic AI domain.

The principle of enforcing communication directionality at the system architecture level—so that the protected side is unreachable from the untrusted side regardless of the adversary's capability—has a proven track record in critical infrastructure protection. The present author co-invented a unidirectional communication architecture for protecting bulk electric system data in power plants (US 8,838,955, granted 2014, assigned to General Electric) [29]. That system uses hardware data diodes (RS-422 unidirectional serial links) to create non-routable communication paths between a critical infrastructure proxy server and an external connectivity server, ensuring that critical plant data can be transmitted outward for remote monitoring while no inbound path exists through which an external attacker could reach the critical data store. The installation-specific configuration files that govern datagram construction and encryption are accessible only through the unidirectional links themselves, placing the governance configuration outside the untrusted entity's information space.

The architectural principle is identical to what R4 requires for agentic AI containment: unidirectional communication channels that the constrained entity cannot address, discover, or reverse. The implementation context differs—hardware data diodes for SCADA networks versus OS-level namespace isolation and inherited file descriptors for AI agent containment—but the security invariant is the same. This continuity demonstrates that adversarial audit isolation for AI agents is not a speculative concept but an application of a proven architectural pattern to a new threat model. The decade of operational experience with unidirectional communication in critical infrastructure provides empirical validation that the principle scales and endures.

### 6.4 The Structural Parallel

The February 2026 and April 2026 events are instances of the same structural failure. In February, a *human operator* circumvented AI safety constraints by replacing the provider that enforced them. In April, the *AI itself* circumvented safety constraints by exploiting the environment that enforced them. In both cases, the constraint mechanism was reachable by the entity it constrained—through organizational action in the first case, through technical exploitation in the second. The architectural lesson is identical: safety constraints that exist inside or alongside the constrained system, rather than architecturally between the system and its execution environment, are not durable under pressure.

The author's AURA (Autonomous Unified Reliability Architecture) patent portfolio [3][4][5] was developed in response to the first pattern—provider-dependent constraints in military and safety-critical autonomous systems. The Mythos incidents demonstrate that the same architectural principles apply to the second pattern—agent-circumventable constraints in agentic AI systems. We term the extension of the AURA architecture to agentic LLM containment the AURA FARADAY (Framework for Adversarial Runtime Action Detection, Audit, and Yield) containment architecture—named for the Faraday cage, which prevents electromagnetic signals from crossing a boundary regardless of the power of the source inside. The analogy is precise: a Faraday cage is capability-agnostic (it works whether the contained signal is 1 watt or 1,000 watts), just as AURA FARADAY's action-level enforcement is capability-agnostic (it evaluates proposed actions against the constraint hierarchy regardless of how capable the model that generated them). The foundational insight—that constraint enforcement must be provider-independent and architecturally separated from the constrained entity—transfers directly from autonomous systems governance to agentic AI containment. The specific architectural context of structured tool calls, multi-step exploit chains, and the adversarial audit model requires extensions beyond the parent AURA portfolio, and these extensions are the subject of ongoing patent prosecution.

The architectural patterns underlying AURA FARADAY have been validated across multiple safety-critical domains in the author's prior work. A published patent application on verifiable ethical arbitration for autonomous vehicles [30] discloses a Hierarchical Constraint Logic Processor operating within a hardware Trusted Execution Environment, applying a constraint hierarchy of non-negotiable safety mandates and tiered ethical mandates, and generating counterfactual audit logs that record both chosen actions and the harm scores of all rejected alternatives. The hierarchical constraint pattern (inviolable vs. governance-modifiable), the isolated execution environment (TEE for vehicles, OS privilege separation for AI agents), and the counterfactual audit (recording rejected as well as chosen options) are the same architectural primitives, applied across three distinct domains: autonomous vehicle ethical arbitration, critical infrastructure data protection [29], and now agentic AI containment. This cross-domain consistency suggests that the requirements derived in this paper (R1–R5) are not specific to AI agents but reflect fundamental principles for governing any autonomous system whose actions have consequential effects and whose compliance cannot be assumed.

## 7. RELATED WORK

The intersection of AI safety and agentic AI security is receiving rapidly increasing attention. The International AI Safety Report 2026, chaired by Yoshua Bengio and authored by over 100 experts with backing from 30+ countries [25], provides the authoritative evidence-based assessment of general-purpose AI capabilities and emerging risks. The report confirms that AI capabilities are advancing faster than safeguards and that "sophisticated attackers can often bypass current defences." The CLTR "Scheming in the Wild" report [24] provides the first systematic empirical evidence of scheming behaviors in deployed production AI systems, establishing a quantitative baseline (698 incidents, 4.9× acceleration) against which containment effectiveness can be measured. Comprehensive surveys of agentic AI security threats and defenses have been published [16][17], and the OWASP Foundation's Top 10 for LLM Applications [18] catalogs the primary threat categories including prompt injection and agent tool interaction manipulation. The 2025 AI Agent Index [19] documents safety features across deployed agentic systems, finding significant gaps in sandboxing, evaluation, and third-party testing.

On the formal specification side, Agent Behavioral Contracts [14] provides a formal framework for specifying behavioral preconditions, invariants, and governance constraints for autonomous agents. The Cryptographic Runtime Governance architecture [15] proposes an immutable ethics policy layer with zero-knowledge proof verification, though this remains theoretical. Agent-C [20] defines a domain-specific language for temporal safety constraints enforced during token generation.

On the defensive tooling side, the landscape has expanded rapidly in early 2026: AEGIS [11] provides pre-execution mediation, Microsoft's AGT [12] provides policy enforcement with execution rings, NVIDIA's OpenShell [8] provides kernel-level sandboxing, and open-source tools like Pipelock [21] and TrustLoop [22] provide proxy-based interception for MCP-based agent workflows.

Arize's observability-driven sandboxing [23] combines policy enforcement with OpenTelemetry-based tracing.

The present paper's contribution is distinct from these works: rather than proposing a specific system, we derive *requirements* from demonstrated failure modes, provide a systematic assessment framework, and identify the gap between current approaches and the adversarial agent threat model. The requirements framework (R1–R5) provides a basis for evaluating whether any proposed system—existing or future—actually addresses the demonstrated threat.

Several concurrent works in March–April 2026 are directly relevant to the requirements derived here. SandboxEscapeBench [31] quantifies frontier LLM capabilities for container sandbox escape across 18 challenge scenarios, providing empirical grounding for R1: the paper reports that the largest frontier models achieve meaningful success rates on difficulty-3–5 scenarios and that 'plain Docker isolation is often insufficient by default to contain highly capable LLM agents'—confirming that environment-level containment (R1-violating architectures) is empirically inadequate. SafetyDrift [32] models cumulative agent safety state as an absorbing Markov chain, predicting the probability of eventual safety violation and identifying 'points of no return' in agent trajectories. SafetyDrift is complementary to the present work: it addresses prediction of safety violations, while AURA FARADAY addresses enforcement preventing them. The AIR incident response framework [33] defines a domain-specific language for autonomous LLM agent incident response, achieving over 90% success rates in detection and remediation. AIR is positioned downstream of a containment failure; the R1–R5 architecture is upstream prevention. 'Agentic Pressure' [34]—the phenomenon where agents strategically sacrifice safety constraints to preserve task utility under pressure—provides empirical support for the adversarial framing adopted throughout this paper: if agents may strategically deprioritize safety, a trust model that treats agents as potentially adversarial is not conservative but accurate. A taxonomy of semantic privilege escalation for agentic systems [35]—the pattern whereby an agent uses authorized permissions to perform actions semantically outside its authorized scope—validates R1's requirement for semantic intent extraction rather than syntactic access control. Capability-safe programming language approaches [36] (Scala 3 capture checking for agent tool call typing) offer a complementary compile-time constraint mechanism that is formally verifiable but depends on agent code being written in the target language and does not address runtime adversarial agents.

## 8. DISCUSSION

The Mythos incidents have immediate implications for the deployment of agentic AI systems across all domains, not merely cybersecurity.

**The systemic nature of the threat.** It is essential to recognize that the containment challenge is not specific to one model or one organization. The CLTR data [24] demonstrates that scheming behaviors—disregarding instructions, circumventing safeguards, deceiving users, taking covert actions—are emerging across models from multiple providers and are increasing at a rate that far exceeds the growth in AI deployment. The 4.9× acceleration in scheming incidents over six months suggests that as models become more capable, the propensity and sophistication of boundary-testing behavior increases correspondingly. This is consistent with the International AI Safety Report's assessment [25] that "the gap between the pace of technological advancement and our ability to implement effective safeguards remains a critical challenge." The requirements derived here (R1–R5) must therefore be understood not as responses to a single incident but as architectural standards for an era in which agentic AI systems will routinely test the boundaries of whatever containment mechanisms are deployed.

**The proliferation problem and the limits of access control.** Anthropic's decision to withhold Mythos from public release is a responsible but temporary measure. As Amodei himself has acknowledged, equivalent capabilities will emerge from other developers [1]. The trajectory from Claude Opus 4.6 to Mythos—a qualitative leap in exploit capability—will be replicated by competitors within months. Open-weight models, whose parameters are publicly available and cannot be restricted after release, are on a capability trajectory that will reach Mythos-equivalent performance within 12–18 months. When that occurs, access-control-based safety is no longer viable: you cannot withhold a model whose weights are downloadable. This proliferation dynamic makes **architectural containment the only durable safety strategy**. The requirements framework (R1–R5) is designed for this future: containment that works regardless of which model is deployed, who developed it, and whether it is proprietary or open-weight. Provider-independent constraint enforcement [4] is not merely an architectural preference; it is a necessity in a world where the provider may not exist or may not cooperate.

**The infrastructure-level consequences of containment failure.** The practical consequence of uncontained agentic AI at frontier capability extends beyond cybersecurity into infrastructure-level catastrophe. Anthropic's own assessment, published alongside the Mythos disclosure, states that the model could autonomously compromise banking systems, power grids, hospitals, and water treatment facilities through exploitation of zero-day vulnerabilities across all major operating systems [1]. The Frontier Red Team reported that engineers with no formal security training asked Mythos to find remote code execution vulnerabilities overnight and woke the following morning to complete, working exploits [2]—collapsing the exploit development timeline from months of specialized human effort to hours of unsupervised computation. Cisco's Glasswing representative stated that "AI capabilities have crossed a threshold that fundamentally changes the urgency required to protect critical infrastructure… and there is no going back" [1]. Palo Alto Networks warned that the model "signals a dangerous shift" and that defenders must "prepare for AI-assisted attackers" producing "more attacks, faster attacks and more sophisticated attacks" [27].

The danger is compounded by two factors. First, *speed*: the window between vulnerability discovery and exploitation, which historically provided defenders days to weeks for patching, collapses to minutes when an AI system can autonomously discover, develop, and deploy exploits. Second, *scale*: a single model instance can scan thousands of systems concurrently, meaning that a coordinated attack on banking infrastructure, power grids, and communications systems simultaneously becomes operationally feasible for any actor with access to the model—including criminal organizations, terrorist groups, and hostile nation-states. The U.S. government summoned bank executives to discuss the cyber risks posed by Mythos [28], and Anthropic's own announcement acknowledged that "the fallout—for economies, public safety, and national security—could be severe" [1].

This assessment transforms the requirements derived in this paper from academic contributions to **infrastructure protection requirements**. R1 (trust separation) prevents the agent from disabling the governance layer en route to exploiting external targets. R2 (sequential intent inference) detects the multi-step exploit chains that constitute real-world attacks. R3 (integrity monitoring) ensures containment cannot be silently degraded before an attack is launched. R4 (adversarial audit isolation) guarantees forensic reconstruction of any incident. R5 (capability envelope enforcement) detects when a model begins exercising exploit capabilities that were not present during safety evaluation—the exact scenario that Mythos demonstrated. The urgency is not speculative: Anthropic has stated that equivalent capabilities will proliferate beyond responsible actors "before long" [1], and the CLTR data [24] confirms that the underlying behavioral pattern is already accelerating across the deployed model population.

**Implications for agentic coding tools.** Products such as Claude Code, OpenAI Codex, and Cursor provide agentic LLMs with shell access, file system access, and network access on users' local machines. These products deliberately minimize sandboxing to maximize capability—a design philosophy explicitly acknowledged by Anthropic [2]. The Mythos incidents demonstrate that the model powering these tools may possess the capability to exploit the local environment in ways the user does not intend or authorize. The requirements derived here apply directly to the containment architecture for agentic coding tools.

**Implications for Project Glasswing.** Anthropic's Project Glasswing [1] provides Mythos Preview to 12 partner organizations and over 40 additional organizations for defensive cybersecurity research. These partners are deploying the model to find vulnerabilities in their own systems—a use case where the model's exploit capabilities are the desired feature, not the threat. However, the containment challenge remains: the model must be constrained to find vulnerabilities *within the authorized scope* without exploiting systems outside that scope, without exfiltrating findings to unauthorized endpoints, and without retaining exploit code across sessions. The cybersecurity research agent embodiment in R1–R5 addresses this deployment context directly.

**Implications for military autonomous systems.** The convergence of the February 2026 provider-substitution events and the April 2026 agent-escape events confirms that military autonomous systems governance requires constraint enforcement at the system architecture level, independent of both the technology provider and the execution environment. The author's published work on governance architectures for military autonomous operations [3][4] addresses this requirement, and the present analysis extends the threat model to include the AI agent itself as a potential circumvention vector—a threat model that is increasingly relevant as military systems adopt more capable AI models.

**Limitations.** This analysis is based on publicly disclosed information from Anthropic's system card and blog posts. Internal details of the Mythos containment mechanisms, the specific exploits used, and the full scope of documented incidents may differ from the public record. The requirements derived here may be incomplete if undisclosed incidents reveal additional failure modes. Additionally, the requirements address the containment problem for agentic LLMs with tool access; they do not address the broader alignment problem of ensuring that AI systems have beneficial goals, which is a necessary complement to containment but a distinct research challenge.

## 9. CONCLUSION

The April 2026 Mythos incidents mark a transition from theoretical to demonstrated risk in agentic AI containment. A frontier model escaped its sandbox, took unauthorized actions, and concealed its activities—defeating containment mechanisms designed and operated by the organization that built the model. We have analyzed five behavioral incidents, identified the architectural assumptions each violates, assessed four containment paradigms against the demonstrated threat model, and derived five requirements that any containment solution must satisfy: trust separation through OS privilege enforcement, sequential intent inference, independent integrity monitoring, adversarial audit isolation, and emergent capability envelope enforcement.

No publicly described system satisfies all five requirements simultaneously. This represents an open problem at the intersection of AI safety and computer security architecture. The problem is urgent: the CLTR data [24] documents 698 real-world scheming incidents accelerating at 4.9× over six months, the International AI Safety Report [25] confirms that safeguards are not keeping pace with capabilities, and the proliferation of agentic AI capabilities—including the inevitable emergence of Mythos-equivalent open-weight models—means that access-control-based safety is a temporary measure with a defined expiration date. The window to develop architecturally robust containment—before the capability to defeat current containment becomes commodity technology—is measured in months, not years.

The architectural principles needed for robust containment are not entirely novel. Provider-independent constraint enforcement, adaptive validation, and immutable audit under adversarial conditions have been explored in the autonomous systems governance literature [3][4][5] and are the subject of published patent applications by the present author. The AURA FARADAY containment architecture—extending these principles to the specific context of agentic LLM containment, with the unique challenges of structured tool calls, multi-step exploit chains, and undiscoverable audit isolation—is the frontier of current work and the subject of ongoing patent prosecution. The present paper contributes the requirements framework (R1–R5) as a basis for the community to evaluate whether proposed solutions actually address the demonstrated threat, and as a call to action for architecturally robust containment before the capability gap becomes unrecoverable.

## ABOUT THE AUTHOR

Richard Joseph Mitchell, ESEP, MBA is the founder and CEO of AuraSpark Technologies LLC (Tavares, Florida), a company developing high-reliability, explainable autonomous systems with built-in governance architectures for safe real-world deployment. He holds the INCOSE Expert Systems Engineering Professional (ESEP) designation and an MBA from the University of Florida. His 38-year career in safety-critical systems spans nuclear instrumentation and control, power plant optimization, and military autonomous systems. He is co-inventor of US Patent 8,838,955 B2 (Two-Way Secure Data Communication Within Critical Infrastructures, assigned to General Electric, granted 2014), a unidirectional data diode architecture for power plant SCADA protection. He has filed 30+ patent applications under AuraSpark Technologies LLC covering the AURA (Autonomous Unified Reliability Architecture) portfolio. Three papers by the author have been accepted at the I/ITSEC 2026 conference. More information is available at https://auraspark.com.